\documentclass[useAMS,usenatbib,usegraphicx]{mn2e}
\usepackage{widetext}
\title[GR polarized radiative transfer]
{General relativistic polarized radiative transfer: \\ building a dynamics-observations interface}   
\author[Shcherbakov and Huang]
{Roman V. Shcherbakov$^{1}$\thanks{E-mail: rshcherbakov@cfa.harvard.edu}\thanks{http://www.cfa.harvard.edu/\%7Ershcherb/} and Lei Huang$^{2},^{3},^{4}$\\
$^{1}$Harvard-Smithsonian Center for Astrophysics, 60 Garden Street, Cambridge, MA 02138, USA\\
$^{2}$Academia Sinica, Institute of Astronomy and Astrophysics, Taipei 106, Taiwan\\
$^{3}$Key Laboratory for Research in Galaxies and Cosmology, Shanghai Astronomical Observatory, Shanghai 200030, China\\
$^{4}$Key Laboratory for Research in Galaxies and Cosmology, The University of Sciences and Technology of China, Hefei 230026, China}

\pagerange{\pageref{firstpage}--\pageref{lastpage}} \pubyear{2010}

\begin{document}
\maketitle
\label{firstpage}

\begin{abstract}
The rising amount of polarized observations of relativistic sources requires the correct theory for proper model fitting. The equations for general relativistic (GR) polarized radiative transfer are derived starting from the Boltzmann equation and basic ideas of general relativity. The derivation is aimed at providing a practical guide to reproducing the synchrotron part of radio \& sub-mm emission from low luminosity active galactic nuclei (LLAGNs), in particular Sgr A*, and jets. The recipe for fast exact calculation of cyclo-synchrotron emissivities, absorptivities, Faraday rotation and conversion coefficients is given for isotropic particle distributions. The multitude of physical effects influencing simulated spectrum is discussed. The application of the prescribed technique is necessary to determine the black hole (BH) spin in LLAGNs, constraining it with all observations of total flux, linear and circular polarization fractions, and electric vector position angle as functions of the observed frequency.
\end{abstract}

\begin{keywords}
 black hole physics -- Galaxy: centre -- plasmas -- polarization -- radiation mechanisms: general -- radiative transfer
\end{keywords}

\section{INTRODUCTION} The good model of radiative transfer is the key in bridging the plasma dynamics and the observations of compact accreting sources. The dynamics of plasma evolved from hydrodynamics \citep{ruffert} to magneto hydrodynamics (MHD) \citep{hawley} and particle-in-cell (PIC) simulations \citep{sironi}. The modelling of compact object's gravity has turned from quasi-Newtonian potential \citep{hawley01,igumenshchev} to the full general relativistic (GR) MHD \citep{villiers,shafee}. Only GRMHD simulations allow to fully account for the spin of the compact object. The radiative transfer approximations were improving as well. The simple quasi-Newtonian ray propagation \citep{CK} gave way to null-geodesics tracing in Kerr metric \citep{schnittman,noble,moscibr_sim}. The huge amount of polarization observations demanded the polarized radiative transfer.

The main principles of GR polarized radiative transfer were formulated in \citet{broderick_thesis}. However, that formulation was not ready for applications as it lacked, for example, the Faraday conversion and the suppression of Faraday rotation in hot plasmas. The first application to the real object was done in \citet{huangnew}. Their calculations included Faraday conversion, but made several approximations, some of which can be substantially improved upon. For example, their simple relation on $V$ and $Q$ emissivities and Faraday rotation and conversion constitutes an approximation that almost never holds. Their emissivities are calculated in synchrotron regime, which breaks for temperatures about the electron mass. Their frame of plasma does not fully account for the fluid motion. We are improving on their work in the present paper, in particular treating exactly the plasma response and extending it to non-thermal particle distributions.

Another important issue is the complexity of GR polarized radiative transfer. The errors and implicit strong approximations may slip into the equations of almost every author. This is more likely the case, when certain derivation is done half-way by one author and then continued by another author, e.g. the derivation of the Faraday conversion coefficient in the mixture of thermal and non-thermal plasmas in \citet{melrose97} and \citet{ballantyne} neglected the importance of the finite ratio of cyclotron to observed frequencies. Another good example is the definition of the sign of circular polarization $V.$ It varies from article to article and the consistent definition in a single derivation is essential.

Therefore, there appeared a need for the present paper. In a single derivation from the basic principles we provide the necessary applied expressions for GR polarized radiative transfer. We start in \S \ref{s_transf} by consistently defining the polarization tensor, Stokes parameters, and plasma response tensor and incorporating the response tensor into the Newtonian radiative transfer. In \S \ref{s_resp} we recast the derivation of the response tensor from Boltzmann equation for general isotropic particle distribution and do the special case of thermal distribution. We provide the applied expressions for response tensor in the plane perpendicular to the ray and give the consistent sign notation for both positive and negative charges in \S \ref{s_rotation}. The resultant formulas for absorptivities/emissivities, Faraday rotation and conversion coefficients are exact and easy to evaluate. By means of locally-flat co-moving reference frame we extend the radiative transfer to full GR in \S \ref{s_gr}. We highlight the various physical effects important for real astrophysical objects in \S \ref{s_appl}. Finally, in \S \ref{s_discus} we briefly summarize the methods and the ways to generalize them even further.
\section{NEWTONIAN POLARIZED RADIATIVE TRANSFER}\label{s_transf}
The proper treatment of polarization of radiation is necessary to take the full advantage of polarized observations.
\begin{figure}
\includegraphics[width=84mm]{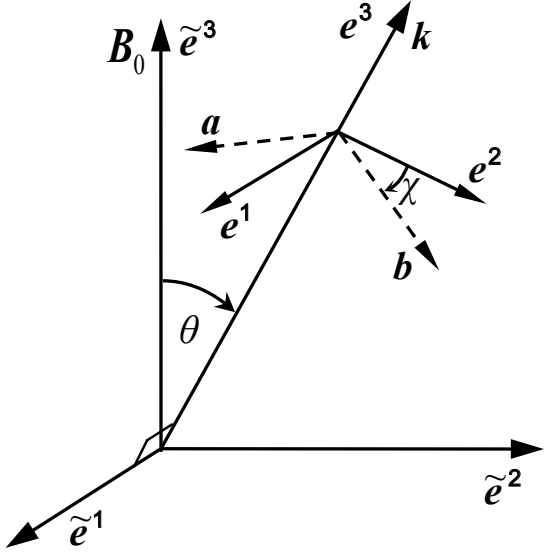}
\caption{Geometry of the problem. Vector ${\bmath B}_0$ represents uniform magnetic field. The transverse plane wave travels along $\bmath k$ and has electric field $\bmath E$ oscillating in $({\bmath e}^1 {\bmath e}^2)$ plane. Vectors $\bmath a$ and $\bmath b$ represent parallel transported basis orthogonalized with $\bmath k.$}\label{fig_vectors}
\end{figure}
Let us start formulating the dynamics of polarization by defining the basis. Let $\tilde{\bmath e}^3$ be the direction of uniform magnetic field ${{\bmath B}_0}.$
Then define the orthonormal triad ${\bmath k},$ ${\bmath e}^1,$ and ${\bmath e}^2$ in the standard way \citep{rybicki,sazonov,pach}:
wave propagates along ${\bmath k}$ vector,
\begin{eqnarray}
{\bmath e}^1&=&C ({\bmath B}_0\times {\bmath k}),\\
{\bmath e}^2&=&{\bmath k}\times {\bmath e}^1,\nonumber
\end{eqnarray} where the scalar $C$ can have either sign. We choose the axes as on Fig.~\ref{fig_vectors}: ${\bmath e}^1$ is perpendicular to $({\bmath B}_0,{\bmath k})$ plane and ${\bmath B}_0$
lies in $({\bmath k},{\bmath e}^2)$ plane. The rotation around ${\bmath e}^1$ transforms basis ($\tilde{\bmath e}^1,$ $\tilde{\bmath e}^2,$ $\tilde{\bmath e}^3$) to the basis
(${\bmath e}^1,$ ${\bmath e}^2,$ ${\bmath e}^3$) as
\begin{eqnarray}
{\bmath e}^1=\tilde{\bmath e}^1,\quad {\bmath e}^2=\tilde{\bmath
e}^2\cos\theta - \tilde{\bmath e}^3  \sin\theta, \\
\quad {\bmath e}^3={\bmath k}=\tilde{\bmath e}^2 \sin\theta + \tilde{\bmath e}^3 \cos\theta,\nonumber
\end{eqnarray}
which can be conveniently written as
\begin{equation}
{\bmath e}^k=\tilde{\bmath e}^i {\mathbfss M}_{i k}, \quad {\mathbfss M}_{i k}=\left(\begin{array}{ccc}
1 & 0 & 0 \\
0 & \cos\theta  & \sin\theta\\
0 &-\sin\theta  & \cos\theta
\end{array}     \right).
\end{equation}
Vectors  and tensors then rotate according to
\begin{equation}\label{transform}
A_k=({\mathbfss M}^T)_{k i}\tilde{A}_i, \quad
\alpha_{k i}=({\mathbfss M}^T)_{k m}\tilde{\alpha}_{m n}{\mathbfss M}_{n i},
\end{equation} where $()^T$ is a transposed matrix and the quantities with tildes are taken in a frame with the basis $(\tilde{\bmath e}^1,$ $\tilde{\bmath e}^2,$ $\tilde{\bmath e}^3$). The angle $\theta$ can be found from
\begin{equation}
\cos\theta=\frac{{\bmath k}\cdot{\bmath B}_0}{k B_0}.
\end{equation}

\begin{figure}
\includegraphics[width=84mm]{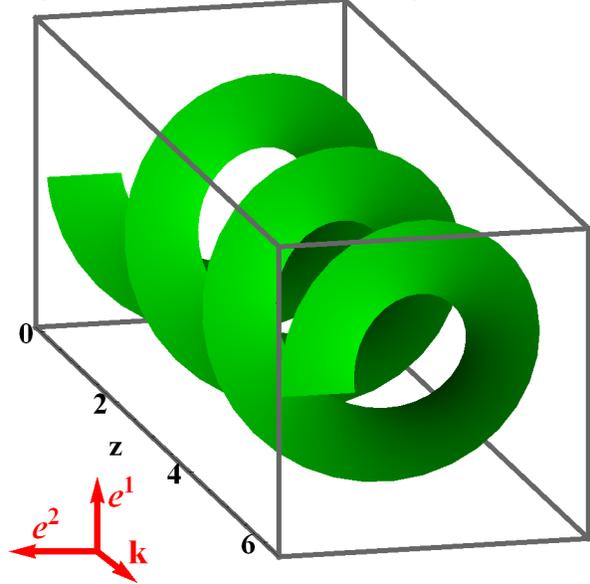}
\caption{Right-handed rotation of electric field along the ray at fixed time $t$ corresponding to negative circular polarized wave $V<0.$
Electric field is $E_1=E_{1\omega} \exp(\imath(k z+\delta)),$ $E_2=E_{2\omega} \exp(\imath k z),$ $\delta=\pi/2,$ where $E_{1\omega}$ is along ${\bmath e}^1,$  $E_{2\omega}$ along ${\bmath e}^2,$ and the wave propagates
along $\bmath k.$ Vectors ${\bmath e}^1,$ ${\bmath e}^2,$ $\bmath k$ constitute the right-handed triad.}\label{figz}
\end{figure}

\begin{figure}
\includegraphics[width=84mm]{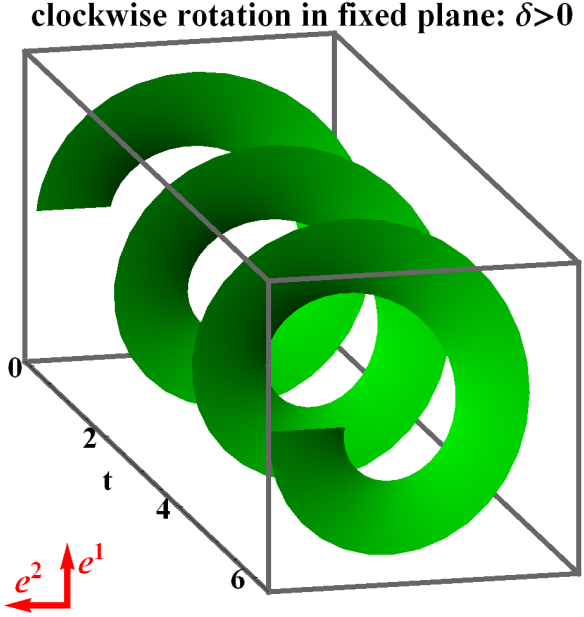}\caption{Left-handed rotation of electric field at fixed coordinate $z$ with time corresponding to negative circular polarized wave $V<0.$
Electric field is $E_1=E_{1\omega} \exp(\imath(-\omega t+\delta)),$ $E_2=E_{2\omega} \exp(-\imath \omega t),$ $\delta=\pi/2,$ where $E_{1\omega}$ is along ${\bmath e}^1,$  $E_{2\omega}$ along ${\bmath e}^2.$}\label{figt}
\end{figure}

For the electric field components $E_1$ along ${\bmath e}^1$ and $E_2$ along ${\bmath e}^2$ the Stokes parameters are defined as
\begin{eqnarray}\label{stokes}
I=\left<E_1 E_1^*\right>+\left<E_2 E_2^*\right>,\nonumber\\
Q=\left<E_1 E_1^*\right>-\left<E_2 E_2^*\right>,\\
U=\left<E_1 E_2^*\right>+\left<E_2 E_1^*\right>,\nonumber\\
V=\imath(\left<E_1 E_2^*\right>-\left<E_2 E_1^*\right>)\nonumber,
\end{eqnarray} where the last formula chooses the IAU/IEEE definition \citep{hamaker} of $V,$ actively used by observers. That is for positive $V$ the rotation of electric field is counter-clockwise
as seen by the observer. Nevertheless, all astrophysics textbooks agree on the opposite definition
of $V$ \citep{sazonov_ts,legg,rybicki,rochford,wilson}. Let us visualize the electric field rotation. Take the electric field
\begin{equation}
E_1=E_{1\omega} \exp(\imath(k z-\omega t+\delta)), \quad E_2=E_{2\omega}\exp(\imath (k z-\omega t))
\end{equation} with positive amplitudes (Fourier coefficients) $E_{1\omega}, E_{2\omega}>0$ and substitute it to the definition (\ref{stokes}). Then
\begin{eqnarray}
I&=&E_{1\omega}^2+E_{2\omega}^2, \qquad Q=E_{1\omega}^2-E_{2\omega}^2\\
U&=&2E_{1\omega} E_{2\omega}\cos\delta, \quad V=-2E_{1\omega} E_{2\omega}\sin\delta\nonumber
\end{eqnarray} and $V<0$ for $\delta\in(0,\pi).$ Let us fix time $t=0,$ $\delta=\pi/2$ and draw the electric field vector in space along the ray (see Fig.~\ref{figz}).
We see that the electric field corresponds to a right-handed screw. However, if we fix a plane in space by setting $z=0$ and draw the evolution of
the electric field vector, then the rotation direction is the opposite: electric field vector rotates counter-clockwise, if viewed along the ray (see Fig.~\ref{figt}).
These opposite directions of rotation is a common point of confusion \citep{rochford}. Correspondingly, the observer sees the clockwise-rotating electric field for $\delta=\pi/2$, as she
is situated at a fixed $z,$ and counter-clockwise rotating electric field for $\delta=-\pi/2$ and positive $V>0.$
The definition (\ref{stokes}) of $V$ has a marginal advantage: the electrons generate positive $V$ for propagation along the magnetic field.
 Let us take an electron on a circular orbit in $(\tilde{\bmath e}^1\tilde{\bmath e}^2)$ plane. It moves from the direction of $\tilde{\bmath e}^1$
 to the direction of $\tilde{\bmath e}^2,$ id est clockwise as viewed along the magnetic field. Then the wake of the electric field follows the charge and rotates clockwise.
 The resultant wave propagates along ${\bmath B}_0$ and constitutes a left-handed screw, which gives the positive $V>0$.

 The polarization tensor is obtained automatically from equation (\ref{stokes}) as
\begin{equation}\label{pol_tensor}
I_{i j}=\left<E_i E_j^*\right>=\frac12
\left(%
\begin{array}{cc}
  I+Q &  U-\imath V\\
  U+\imath V & I-Q \\
\end{array}%
\right).
\end{equation} The polarization vector is
\begin{equation}\label{pol_vector}
{\bmath S}=(I,Q,U,V)^T.
\end{equation} Note that \citet{melrose_dispersive} uses the same definitions, however, their ${\bmath k},$ ${\bmath e}^1,$ ${\bmath e}^2$ constitute a left triad instead
of a right triad.

The plasma response is characterized by the $4\times4$ response tensor ${\alpha^\mu}_\nu$
\begin{equation}\label{response_def}j^\mu_\omega={\alpha^\mu}_\nu
A^{\nu}_\omega\end{equation} as the proportionality between the four-vectors of vector potential amplitude and the current density amplitude. There is freedom in choosing the gauge condition for $A^\mu.$ Let us choose the Lorenz gauge
\begin{equation}\label{gauge}
A^\mu u_\mu=0
\end{equation} at each point along the ray and enforce it by adding to $A^\mu$ a vector, proportional to $k^\mu,$ what does not change the polarization tensor (\ref{pol_tensor}) \citep{misner}.
The gauge condition makes $A^0=\phi=0$ and establishes the proportionality of wave electric field $\bmath E$ and $\bmath A$ in the locally flat co-moving frame \citep{anile,landau2}. Thus, in that frame the spatial components of ${\alpha^\mu}_\nu$ coincide exactly with the spatial $3\times3$ response tensor $\alpha_{ik}$ in $(j_\omega)_i=\alpha_{ik}(A_\omega)_k.$ We will derive the tensor $\alpha_{ik}$ below. It is usually incorporated
within the dielectric tensor
\begin{equation}\label{dielectric}
\varepsilon_{ik}=\delta_{ik}+\frac{4\pi c}{\omega^2}\alpha_{ik}.
\end{equation}
  The wave equation for transverse waves in terms of
$\varepsilon_{ik}$ is
\begin{equation}\label{dispersion}
(n_{\rm r}^2 \delta_{ik} - \varepsilon_{ik}) \left(\begin{array}{c} E_1 \\ E_2
\end{array}\right)=0,
\end{equation} where the indices $i,k=1,2,$ so that only the transverse $2\times2$ part of
$\varepsilon_{ik}$ in $({\bmath e}^1,$ ${\bmath e}^2,$ ${\bmath k})$ basis is taken, and
$n_{\rm r}^2=k^2 c^2/\omega^2$ \citep{swanson,shcher_farad}. The correspondent transport equation is \citep{melrose_dispersive}
\begin{equation}\label{transport_E}
\frac{dE_i}{ds}=\frac{\imath \omega}{2n_{\rm r}c}\Delta{\varepsilon_{ik}} E_k,
\end{equation} where $\Delta{\varepsilon_{ik}}=\varepsilon_{ik}-\delta_{ik}.$
We take the equation (\ref{transport_E}), its conjugate, multiply correspondingly by $E^*_k$ and $E_k,$ add and obtain
\begin{equation}\label{transport_EE}
\frac{d\left<E_i E^*_k\right>}{ds}=\frac{\imath}{\nu}(\alpha_{il}\left<E_l E^*_k\right>-\alpha^*_{kl}
\left<E_i E^*_l\right>)
\end{equation} for $n_{\rm r}=1$ with the observed frequency $\nu=\omega/(2\pi)$ neglecting emission. Again, $i,k,l=1,2.$ Note that $\alpha_{12}=-\alpha_{21}$ according to Onsager principle \citep{landau10} (p. 273). By solving
the definitions of the Stokes parameters (eq. (\ref{stokes})) for $\left<E_i E^*_k\right>$ and substituting the result into the equation
(\ref{transport_EE}) we get the transport equation for the Stokes parameters
\begin{equation}\label{transfer}
\frac{d{\bmath S}}{ds}
=
\left(\begin{array}{c}
  \varepsilon_I \\  \varepsilon_Q \\  0 \\  \varepsilon_V
\end{array}\right)-
\left(%
\begin{array}{cccc}
  \alpha_I & \alpha_Q & 0 & \alpha_V \\
  \alpha_Q & \alpha_I & \rho_V & 0 \\
  0 & -\rho_V & \alpha_I & \rho_Q \\
  \alpha_V & 0 & -\rho_Q & \alpha_I \\
\end{array}%
\right)\bmath S
\end{equation} with the polarization $\bmath S$ vector (\ref{pol_vector}) by adding the emission, where
\begin{eqnarray}\label{ar_def}
\alpha_I&=&{\rm Im}(\alpha_{22}+\alpha_{11})/\nu,\nonumber\\
\alpha_Q&=&{\rm Im}(\alpha_{11}-\alpha_{22})/\nu,\nonumber\\
\alpha_V&=&2{\rm Re}(\alpha_{12})/\nu,\\
\rho_V&=&2{\rm Im}(\alpha_{12})/\nu,\nonumber\\
\rho_Q&=&{\rm Re}(\alpha_{22}-\alpha_{11})/\nu.\nonumber
\end{eqnarray}

\section{DERIVATION OF RESPONSE TENSOR}\label{s_resp}
The response tensor is derived for thermal plasma in \citet{trubnikov,melrose97,swanson}. However, there is a need to recast the derivation, since
we want to consider both signs of charge, extend the results to non-thermal isotropic particle distributions, and seek for extensions to non-isotropic distributions. The importance of notation consistency cannot be overemphasized, also the orientation of our coordinate axes is different from some of the above sources. On the way we discover the practical significance of the response tensor: it offers expressions for fast evaluation of plasma absorptivities and rotativities.

\subsection{General isotropic particle distribution}
Throughout this subsection and next subsection we employ vectors and tensors in ($\tilde{\bmath e}^1,$ $\tilde{\bmath e}^2,$ $\tilde{\bmath e}^3$) basis, but skip tildes to avoid clutter. Only tildes over the response tensors are drawn. For simplicity of notation we define the dimensionless momentum
\begin{equation}
p=\sqrt{\gamma^2-1},
\end{equation} where $\gamma=En/(m c^2)\ge1$ is the $\gamma$-factor for particles of energy $En$ and mass $m.$ Then the Boltzmann equation on the distribution function $f(\bmath x,\bmath p)$ is
\begin{equation}\label{boltzman}
\frac{\partial f}{\partial t}+{\bmath v}\cdot{\bmath \nabla}f+\eta e\left({\bmath E}+
\frac{\bmath v}{c}\times({{\bmath B}_0}+{\bmath B})\right)\cdot\frac{{\bmath \nabla_p}f}{m c}=0.
\end{equation} Here the velocity vector is
\begin{equation}
{\bmath v}={\bmath p}c/\gamma,
\end{equation} ${\bmath B}_0$ is the background magnetic field, 
 ${\bmath E}$ and ${\bmath B}$ are the wave electric and magnetic fields,
$\eta$ is the sign of the charge and $e>0.$ Let us assume a general isotropic particle distribution $f_0(p)$ instead of thermal.
The wave with the phase $\exp(\imath({\bmath k\cdot \bmath r}-\omega t))$ causes perturbation in a form
\begin{equation}\label{f1}
f_1=\exp(\imath({\bmath k\cdot \bmath r}-\omega t))f_0\Phi({\bmath p}),
\end{equation} which implements the general function $\Phi({\bmath p})$ of momentum $\bmath p.$ Note that the perturbation (\ref{f1}) is small and
further analysis is valid only when $|{\bmath B}|\ll |{\bmath B}_0|,$ which corresponds to low radiation pressure medium. Following \citet{trubnikov,swanson} we introduce the cylindrical coordinates with axis along ${\bmath B}_0,$ so that
\begin{equation}
p_x=p_\perp\cos\phi, \quad p_y=p_\perp\sin\phi
\end{equation} with angle  $\phi$ in (xy) plane. Then
\begin{equation}
{\bmath k}\cdot{\bmath p}=k_z p_z+k_\perp p_\perp\sin\phi.
\end{equation} 
The Boltzmann equation (\ref{boltzman}) results in
\begin{eqnarray}
-\imath \omega f_1&+&\frac{\imath(k_z p_z+k_\perp p_\perp \sin\phi)c}{\gamma}f_1\nonumber\\
&+& \eta e {\bmath E}\cdot\frac{{\bmath \nabla_p}f_0}{m c}+\frac{\eta e}{\gamma}
({\bmath p}\times{{\bmath B}_0})\cdot\frac{{\bmath \nabla_p}f_1}{m c}=0
\end{eqnarray} after dropping the second-order terms upon substitution of $f_1$ from equation (\ref{f1}). For general isotropic $f_0(p)$ the relation $({\bmath p}\times({{\bmath B}_0}+{\bmath B}))\cdot{\bmath \nabla_p} f_0=0$ holds, and it does not hold for non-isotropic distributions. Only the cylindrical $\phi$ component
is non-zero in a triple product $({\bmath p}\times{{\bmath B}_0})\cdot{\bmath \nabla_p}f_1.$ After some transformations
we obtain an equation on $\Phi({\bmath p})$
\begin{eqnarray}\label{phi_eq}
-\imath\omega\Phi&+&\frac{\imath(k_z p_z+k_\perp p_\perp\sin\phi)c}{\gamma}\Phi\nonumber\\
&+&\frac{d \ln f_0}{d \gamma}\frac{\eta e}{\gamma m c}({\bmath p}\cdot{\bmath E}_\omega)-\frac{\eta \omega_B}{\gamma}
\frac{\partial \Phi}{\partial \phi}=0
\end{eqnarray} for general isotropic particle distribution, where ${\bmath E}_\omega$ is an amplitude such that
$\bmath E={\bmath E}_\omega\exp(\imath({\bmath k\cdot \bmath r}-\omega t)).$
Here the cyclotron frequency is
\begin{equation}
\omega_B=\frac{e B_0}{m c},\quad \nu_B=\frac{\omega_B}{2\pi}.
\end{equation}  We define the ratio $\beta$
\begin{equation}
\beta=\frac{\nu_B}{\nu}.
\end{equation} Alternatively, the equation (\ref{phi_eq}) reads
\begin{equation}
\imath(a-b\sin\phi)\Phi+\partial \Phi/\partial \phi=F
\end{equation} with
\begin{equation}
a=\frac{\gamma-n_z p_z}{\eta \beta}, \quad b=\frac{n_\perp p_\perp}{\eta \beta}, \quad
F=\frac{d \ln f_0}{d \gamma}{}\frac{{\bmath p}\cdot{\bmath E}_\omega}{B_0}.
\end{equation} Here $n_z=\cos\theta$ and $n_\perp=\sin\theta$ assuming $|{\bmath k}|c=\omega.$
The solution is
\begin{equation}\label{phi_sol}
\Phi(\phi)=\exp(-\imath(a\phi+b\cos\phi))\int^\phi_{\phi_0}\exp(\imath(a\psi+b\cos\psi))F(\psi)d\psi,
\end{equation} where the lower boundary $\phi_0$ is chosen at $t=-\infty.$ The negative charge
moves in the positive $\phi$ direction and the positive charge in the negative $\phi$
direction, therefore $\phi_0=-\eta\infty.$ The solution of a homogeneous equation vanishes over the finite time.
Knowing the particle distribution and the definition of a current density  ${\bmath j}=\eta e \int f_1 {\bmath v}d^3p$
we calculate the current density amplitude
\begin{equation}\label{cur_density}
{\bmath j}_\omega=\eta e \int f_0 \Phi({\bmath p}) \frac{{\bmath p}c}{\gamma}d^3p.
\end{equation} Then we relate it to the electric field ${\bmath E}_\omega$ and the vector potential
${\bmath A}_\omega$ wave amplitudes as
\begin{equation}
(j_\omega)_i=\sigma_{ik} (E_\omega)_k=\alpha_{ik} (A_\omega)_k,
\end{equation} where $\alpha_{ik}=\imath\omega \sigma_{ik}/c.$
Let us calculate $\alpha_{ik}.$ Substituting the solution (\ref{phi_sol}) into the
definition of the current density equation (\ref{cur_density}) and changing the integration variable as
$\psi=\phi-\xi$ we get
\begin{eqnarray}
(j_\omega)_i&=&\imath\frac{\eta e\omega}{c}\int d^3p\int^{-\eta\infty}_0 \exp(-\imath a\xi+\imath b
(\cos(\phi-\xi)-\cos\phi))\nonumber\\
\times&p_i& \left(p_k (A_\omega)_k\right)_{\phi\rightarrow\phi-\xi}\frac{c}{\gamma B_0}\frac{d f_0}{d\gamma}d\xi.
\end{eqnarray} One can
\citep{trubnikov,landau10,swanson} introduce the differentiation with respect to vectors
$\bmath s$ and $\bmath s'$ to eliminate the momenta $\bmath p$ in the integral expressions, then due
to uniform convergence of the integrals in $d\xi$ and $dp$ move the derivatives outside the integrals.
We also do the transformation
$\xi\rightarrow -\eta\beta\xi$ to finally get in ($\tilde{\bmath e}^1,$ $\tilde{\bmath e}^2,$ $\tilde{\bmath e}^3$) basis
\begin{eqnarray}\label{resp_def}
\tilde{\alpha}_{ik}&=&-\frac{\imath e^2}{m c}
\int d^3p\int^\infty_0 \frac{\partial^2 \exp(\imath\xi\gamma-
\imath {\bmath h}\cdot{\bmath p})}{\partial s_i\partial s'_k}\frac{d f_0}{\gamma d\gamma}d\xi \\
&=&-\frac{4\pi\imath e^2}{m c}\frac{\partial^2}{\partial s_i\partial s'_k} \int^\infty_1 d\gamma\frac{d f_0}{d\gamma}\int^\infty_0 \exp(\imath\xi\gamma)\frac{\sin(h p)}h d\xi \nonumber
\end{eqnarray} with
\begin{eqnarray}\label{Ah_def}
h_x&=&\frac{n_\perp}{\beta\eta}(1-\cos(\beta\xi))+\imath(s_x+\cos(\beta\xi)s'_x+\eta\sin(\beta\xi)s'_y),\nonumber\\
h_y&=&\frac{\sin(\beta\xi)n_\perp}\beta+\imath(s_y-\eta\sin(\beta\xi)s'_x+\cos(\beta\xi)s'_y),\nonumber\\
h_z&=&\xi n_z +\imath(s_z+s'_z),\\
h&=&\sqrt{h_x^2+h_y^2+h_z^2}.\nonumber
\end{eqnarray}
\subsection{Thermal particle distribution}
Let us now consider the special case of an isotropic thermal distribution of particles
\begin{equation}\label{distr_T}
f_0=\frac{n_e}{4\pi}\frac{\exp(-\gamma/\theta_e)}{\theta_e K_2(\theta_e^{-1})}
\end{equation} normalized as
\begin{equation}
\int^{+\infty}_0f_0 4\pi p^2 dp=n_e.
\end{equation} Here and below $K_n(x)$ is a Bessel function of the second type of order $n$ with argument $x.$ The normalized particle temperature is
\begin{equation}
\theta_e=\frac{k_B T}{m c^2}.
\end{equation} Then the response tensor (\ref{resp_def}) is
\begin{equation}
\tilde{\alpha}_{ik}=\frac{\imath e^2 n_e}{m c}
\int \int^\infty_0 \frac{\partial^2}{\partial s_i\partial s'_k} \frac{\exp(-A'\gamma-
\imath {\bmath h}\cdot{\bmath p})}{4\pi\gamma \theta_e^2 K_2(\theta_e^{-1})}d\xi d^3p
\end{equation} with $A'=1/\theta_e-\imath\xi$ and the rest of quantities defined by equation (\ref{Ah_def}). The integral over $d^3p$ can be taken analytically \citep{trubnikov} to give
\begin{equation}
\tilde{\alpha}_{ik}=\frac{\imath e^2 n_e}{m c\theta_e^2 K_2(\theta_e^{-1})}
\frac{\partial^2}{\partial s_i \partial s'_k}\int^\infty_0 \frac{K_1(\sqrt{A'^2+h^2})}{\sqrt{A'^2+h^2}}d\xi.
\end{equation} Performing the differentiation in Mathematica 7 to avoid errors, one gets
$3\times3$ response tensor
\begin{equation}\label{alpha33}
\tilde{\alpha}_{ik}=\frac{\imath e^2 n_e}{m c\theta_e^2 K_2(\theta_e^{-1})}\int^\infty_0 d\xi\left(
\tilde{T}^1_{ik}\frac{K_2({\mathcal R})}{{\mathcal R}^2}-\tilde{T}^2_{ik}\frac{K_3({\mathcal R})}{\beta^2{\mathcal R}^3}
\right)
\end{equation} with
\begin{equation}\label{rexp}
{\mathcal R}=\sqrt{\frac{1}{\theta_e^2}-\frac{2\imath\xi}{\theta_e}-\xi^2\sin^2\theta+\frac{2\sin^2\theta}{\beta^2}(1-\cos\beta\xi)}.
\end{equation} Here
\begin{equation}\label{T1}
\tilde{T}^1_{ik}=\left(
\begin{array}{ccc}
  \cos\beta\xi & \eta\sin\beta\xi & 0 \\
-\eta\sin\beta\xi & \cos\beta\xi & 0 \\
0 & 0 & 1 \\
\end{array}
\right),
\end{equation}

\begin{widetext}
\begin{equation}\label{T2}
\tilde{T}^2_{ik}=\left(
\begin{array}{ccc}
-(1-\cos\beta\xi)^2\sin^2\theta & \quad \eta(1-\cos\beta\xi)\sin\beta\xi \sin^2\theta & \quad \eta\beta\xi(1-\cos\beta\xi)\cos\theta\sin\theta \\
-\eta(1-\cos\beta\xi)\sin\beta\xi \sin^2\theta &\sin^2\beta\xi\sin^2\theta & \beta\xi\sin\beta\xi\sin\theta\cos\theta\\
-\eta\beta\xi(1-\cos\beta\xi)\cos\theta\sin\theta  & \beta\xi\sin\beta\xi\sin\theta\cos\theta & \quad\beta^2\xi^2\cos^2\theta \\
\end{array}
\right).
\end{equation}
\end{widetext} The expressions (\ref{alpha33}-\ref{T2}) are hiding inside two almost transverse and one almost longitudinal damped eigenwaves.

\subsection{Rotation of thermal response tensor}\label{s_rotation}
Let us apply the transformation (\ref{transform}) to tensors $\tilde{T}^1_{ik}$ and $\tilde{T}^2_{ik}$ and take the transverse
$2\times2$ part to obtain correspondingly
\begin{equation}
T^1_{ij}=\left(%
\begin{array}{cc}
  \cos\beta\xi & \eta\sin\beta\xi\cos\theta  \\
-\eta\sin\beta\xi\cos\theta & \cos\beta\xi \cos^2\theta+\sin^2\theta\\
\end{array}%
\right)
\end{equation} and $T^2_{ij}=R_i\bar{R}_j$ with
\begin{eqnarray}
R_i=\sin\theta(\eta(1-\cos\beta\xi),\cos\theta(\sin\beta\xi-\beta\xi)),\quad\\
\bar{R}_j=\sin\theta(-\eta(1-\cos\beta\xi),\cos\theta(\sin\beta\xi-\beta\xi))\nonumber
\end{eqnarray} for
\begin{equation}\label{resp_calc}
\alpha_{ij}=\frac{\imath e^2 n_e }{m c\theta_e^2 K_2(\theta_e^{-1})}\int^\infty_0 d\xi\left(
T^1_{ij}\frac{K_2({\mathcal R})}{{\mathcal R}^2}-T^2_{ij}\frac{K_3({\mathcal R})}{\beta^2{\mathcal R}^3}
\right).
\end{equation} The integration over $\xi$ converges very slowly, if performed along the real axis. The way to accelerate the convergence is to perform the integration in a complex plane at a positive angle to the real axis. The wave frequency $\nu$ in the above calculations has a small positive imaginary part ${\rm Im}(\nu)>0$ to account for the energy pumped into particles from passing waves. Then ${\rm Im}(\beta)<0$ and the expression (\ref{rexp}) has zeros only in the lower plane ${\rm Im}(\xi)<0$ of $\xi.$ Thus, deforming the integration contour to the upper plane of $\xi$ does not change the response tensor (\ref{resp_calc}). Note that all absorptivities $\alpha_I,$  $\alpha_Q,$ and $\alpha_V$ and rotativities $\rho_Q$ and $\rho_V$ are positive for electrons for $\theta\in(0,\pi/2)$ under the definitions (\ref{ar_def}), what gives an easy way to check the implementation of radiative transfer algorithm. The evaluation of these coefficients will be reported in the subsequent paper (\citet{huang_shcher}, in prep.). We will also evaluate the validity of a transverse approximation for waves.

Following \citet{huangnew} we define the parallel transported vectors $\bmath a$ and $\bmath b$ in addition to the right triad ${\bmath e}^1,$ ${\bmath e}^2,$ $\bmath k,$
so that in the co-moving locally-flat reference frame
\begin{equation}
({\bmath a},{\bmath b})=({\bmath e}^1,{\bmath e}^2)\left(%
\begin{array}{cc}
  \cos\chi & \sin\chi\\
  -\sin\chi & \cos\chi \\
\end{array}%
\right).
\end{equation} Then the transformation with $-2\chi$ angle
\begin{equation}\label{rmatrix}
{\mathbfss R}(\chi)=\left(%
\begin{array}{cccc}
  1 & 0 & 0 & 0\\
  0 &\cos(2\chi) & -\sin(2\chi) & 0 \\
  0 &\sin(2\chi) & \cos(2\chi) & 0\\
  0 & 0 & 0& 1\\
\end{array}%
\right)
\end{equation} serves to get the vector of emissivities $\bmath \varepsilon$ and the matrix of rotativities/absorptivities $K$ in (${\bmath a},$ ${\bmath b},$ $\bmath k$) basis as
\begin{eqnarray}{\bmath \varepsilon}&=&{\mathbfss R}(\chi)\left(\begin{array}{c}
  \varepsilon_I \\  \varepsilon_Q \\  0 \\  \varepsilon_V
\end{array}\right),\\
{\mathbfss K}&=&{\mathbfss R}(\chi)\left(%
\begin{array}{cccc}
  \alpha_I & \alpha_Q & 0 & \alpha_V \\
  \alpha_Q & \alpha_I & \rho_V & 0 \\
  0 & -\rho_V & \alpha_I & \rho_Q \\
  \alpha_V & 0 & -\rho_Q & \alpha_I \\
\end{array}%
\right){\mathbfss R}(-\chi).\nonumber
\end{eqnarray}
Define the perpendicular magnetic field
\begin{equation}\label{Bperp}
{\bmath B}_{0\perp}={{\bmath B}_0}-{\bmath k}({\bmath k}\cdot{\bmath B}_0)/k^2.
\end{equation}
The trigonometric factors are related to the magnetic field as
\begin{equation}\label{chiang}
\sin\chi=({\bmath a}\cdot{\bmath B}_{0\perp})/B_{0\perp}, \quad \cos\chi=-({\bmath b}\cdot{\bmath B}_{0\perp})/B_{0\perp}.
\end{equation} The radiative transfer equation is then
\begin{equation}\label{tran_main}
d{\bmath S}/ds={\bmath \varepsilon}-{\mathbfss K}{\bmath S}
\end{equation} for the polarization vector $\bmath S$ defined in (${\bmath a},$ ${\bmath b},$ $\bmath k$) basis.

\section{EXTENSION TO GENERAL RELATIVITY}\label{s_gr}
Let us consider two reference frames: locally-flat co-moving reference frame with 4-velocity $\hat{u}^\alpha=(1,0,0,0)$ and flat metric and
the lab frame with Kerr metric and the fluid moving at $u^\alpha.$  We denote by hats $\hat{()}$ the quantities in the co-moving frame.
Consider the radiative transfer equation (\ref{tran_main}) in the co-moving frame. The set of Stokes parameters $\bmath S$ can be generalized to the corresponding set of
photon occupation numbers
\begin{equation}{\bmath N}={\bmath S}/\nu^3,
\end{equation} which are invariant under the orthogonal coordinate transformations \citep{misner,anile,ellis}.
Photons propagate along null-geodesics with the affine parameter $\lambda,$ so that the wave four-vector is
\begin{equation}\label{kalpha}
k^\alpha=k_0\frac{dx^\alpha}{d\lambda},
\end{equation} and
\begin{equation}
\frac{d{\bmath N}}{d\lambda}=0.
\end{equation} Here $k_0$ is a constant photon energy, which relates to the observed frequency as
\begin{equation}
\nu_\infty=\frac{k_0 c}{2\pi}.
\end{equation} Under such normalization
of $\lambda$ approximately $ds\approx d\lambda$ far from the BH. One calculates the null geodesic starting from the observer's plane. The perpendicular unit vector $a^\alpha$ has a special orientation on that plane, and is transported along the geodesic according to
\begin{equation}\label{aalpha}
a^\alpha(\lambda=0)=a^\alpha_0,\quad a^\alpha_0 a_{\alpha0}=1,\quad k^\sigma \nabla_\sigma a^\alpha=0,
\end{equation} where $\nabla_\sigma$ is the covariant derivative. The unit vector $b^\alpha$ is transported the same way.

Just as in a flat space case the charged particles lead to the increase of occupation numbers $\bmath N$ due to emission,
to decrease of $\bmath N$ due to absorption, and to exchange of $\bmath N$ components due to Faraday rotation and Faraday conversion. These are all
the processes occuring in linear regime. The invariant number of photons emitted per unit solid angle per unit frequency per unit volume per unit
 time is proportional to the invariant $\varepsilon(\nu)/\nu^2=\hat{\varepsilon}(\hat{\nu})/\hat{\nu}^2$ \citep{mihalas} as
\begin{equation}
\frac{d{\bmath N}}{d\lambda}\propto \frac{\varepsilon(\nu)}{\nu^2},
\end{equation} where
\begin{equation}
\nu=-k^\mu u_\mu
\end{equation} is the photon frequency in the lab frame for $(-,+,+,+)$ signature of metric. By $\hat{\nu}$ the photon frequency in the co-moving frame is denoted. The invariant change
 of photon states due to absorption and propagation effects is proportional to the co-moving frame matrix ${\mathbfss K}$ (see eq.(\ref{tran_main})) taken within unit frequency unit solid
 angle unit volume unit time. Thus the proportionality to the invariant $\nu {\mathbfss K}(\nu)=\hat{\nu}\hat{\mathbfss K}(\hat{\nu})$ is established  \citep{mihalas} as
\begin{equation}
\frac{d{\bmath N}}{d\lambda}\propto -\nu {\mathbfss K}(\nu) {\bmath N},
\end{equation} where the whole form of the absorption/state change matrix is preserved. The full GR radiative transfer equation
\begin{equation}\label{fin_trans}
\nu_\infty\frac{d{\bmath N}}{d\lambda}=\frac{\hat{\varepsilon}(\hat{\nu})}{\hat{\nu}^2}-\hat{\nu} \hat{\mathbfss K}(\hat{\nu}) {\bmath N}
\end{equation} is obtained. The equation (\ref{fin_trans}) is similar to the GR polarized transfer equation in \citet{huangnew}. However, their usage of primed and unprimed quantities is potentially confusing.
It is their primed quantity ${\bmath S}',$ which should be generalized to GR as ${\mathcal {\bmath N}}_S={\bmath S}'/\nu^3.$

\subsection{Transformation to locally-flat co-moving frame}
The angle $\chi$ in the expression (\ref{chiang}), $\theta$ in the response tensor, and similar quantities need to either be evaluated in the locally-flat reference frame, where fluid is at rest, or properly calculated in GR. We choose the first path as a transparent one with the following recipe. First, one traces the null geodesic from the observer's plane to the BH horizon or the sphere far from the BH and finds the vectors $k^\alpha$ and $a^\alpha$ (see eqs.(\ref{kalpha},\ref{aalpha})). At each point on the ray one knows the vectors $k^\alpha,$ $a^\alpha,$ fluid four-velocity $u^\alpha,$ and the four-vector of magnetic field $B^\alpha_0$ defined in \citet{mckinney} in the lab frame. The next step is to transform all vectors to the co-moving frame. Let us construct an orthonormal basis in Kerr metric in lab frame
\begin{eqnarray}\label{kerr_basis}
e^\alpha_{(t)}=(u^t,u^r,u^\theta,u^\phi),\nonumber\\
e^\alpha_{(r)}\propto(u^tu_r,-(u^tu_t+u^\phi u_\phi),0,u^\phi u_r),\\
e^\alpha_{(\theta)}\propto (u^t u_\theta,u^r u_\theta, u^\theta u_\theta +1,u^\phi u_\theta),\nonumber\\
e^\alpha_{(\phi)}\propto (-u_\phi/u_t,0,0,1),\nonumber
\end{eqnarray} where lower-index velocity is $u_\alpha=g_{\alpha\beta}u^\beta$ and $g_{\alpha\beta}$ is the lower index Kerr metric
\begin{equation}
g_{\alpha\beta}=\left(%
\begin{array}{cccc}
  -1+\frac{2r}{\rho} & 0& 0 & -\frac{2 a r \sin^2\theta_a}{\rho} \\
  0 & \rho/\Delta & 0 & 0 \\
  0 & 0 & \rho & 0 \\
  -\frac{2 a r \sin^2\theta_a}{\rho} & 0 & 0 & \frac{\Sigma \sin^2\theta_a}{\rho} \\
\end{array}%
\right)
\end{equation}  in $(t,r,\theta_a,\phi)$ spherical polar coordinates with polar angle $\theta_a,$ radius $r,$ spin $a,$ $\rho=r^2+a^2\cos^2\theta_a,$ $\Delta=r^2-2r+a^2,$ $\Sigma=(r^2+a^2)^2-a^2\Delta\sin^2\theta_a.$ Then make a transformation to
\begin{eqnarray}\label{hat_basis}
\hat{e}^\alpha_{(t)}=(1,0,0,0),\nonumber\\
\hat{e}^\alpha_{(r)}=(0,1,0,0),\\
\hat{e}^\alpha_{(\theta)}=(0,0,1,0),\nonumber\\
\hat{e}^\alpha_{(\phi)}=(0,0,0,1)\nonumber
\end{eqnarray} via
\begin{equation}\label{local_frame}
S_{(t,r,\theta,\phi)\beta}=(-e^\alpha_{(t)},e^\alpha_{(r)},e^\alpha_{(\theta)},e^\alpha_{(\phi)}) g_{\alpha\beta}.
\end{equation} The transformation of a four-vector $A^\beta$ to the co-moving frame is then
\begin{equation}\label{vect_trans}
\hat{A}_{(t,r,\theta,\phi)}=S_{(t,r,\theta,\phi)\beta}A^\beta.
\end{equation} The metric in the new frame
\begin{equation}
S_{(i)\alpha}g^{\alpha\beta}(S_{(k)\beta})^T=\eta_{ik}
\end{equation} coincides with Minkowski metric $\eta_{ik}={\rm diag}(-1,1,1,1).$ The velocity four-vector $u^\beta$ transforms to $\hat{u}_{(i)}=(1,0,0,0)^T.$ Thus, the basis change (\ref{vect_trans}) with matrix (\ref{local_frame}) and vectors (\ref{kerr_basis}) constitutes the transformation to the locally-flat co-moving reference frame. This procedure is the alternative of the numerical Gramm-Schmidt orthonormalization applied in \citet{moscibr_sim}. The basis vectors (\ref{kerr_basis}) are presented in \citet{krolik,beckwith}, our expressions being a simplified version of vectors in \citet{beckwith}. Note, that despite the vectors (\ref{kerr_basis}) do not explicitly depend on the metric elements, the expressions rely on the properties of Kerr metric and are not valid for general $g^{\alpha\beta}.$

Upon transforming $u^\alpha,$ $k^\alpha,$ $a^\alpha,$ $B^\alpha_0$ to the co-moving frame we easily find the wave frequency $\nu=-\hat{k}_{(0)},$ then $\hat{{\bmath k}}=\hat{k}_{(1,2,3)}.$ The perpendicular vector $\hat{\bmath a}$ needs to be offset by $\hat{\bmath k}$ as
\begin{equation}\label{a_cm}
\hat{\bmath a}=\hat{a}_{(1,2,3)}-\hat{\bmath k}\frac{\hat{a}_{(0)}}{\hat{k}_{(0)}}
\end{equation} and then normalized to construct a spatial unit vector. The offset is due to the enforcement of Lorenz gauge (\ref{gauge}). It conveniently makes ${\bmath a}\cdot{\bmath k}=\hat{\bmath a}\cdot\hat{\bmath k}=0.$ The vector $\bmath b$ is found by a simple vector product
\begin{equation}
\hat{{\bmath b}}=\hat{\bmath a}\times\hat{\bmath k}
\end{equation} and then normalized.  The spatial part ($\hat{\bmath e}_{(r)},$ $\hat{\bmath e}_{(\theta)},$ $\hat{\bmath e}_{(\phi)}$) of basis (\ref{hat_basis}) relates to basis (${\bmath e}^1,$ ${\bmath e}^2,$ ${\bmath e}^3$) via the orthonormal transformation preserving angles. The magnetic field $B^\alpha_0$ gets transformed to a three-vector $\hat{\bmath B}_0$ and all the angles are found in correspondence to Fig.~\ref{fig_vectors} with the help of equations (\ref{Bperp},\ref{chiang}) applied to hatted vectors. For example, $\cos\theta=(\hat{\bmath k}\cdot\hat{\bmath B}_0)/({\hat{k} \hat{B}_0}).$ Then the whole matrix $\hat{\mathbfss K}$ is found.

\section{APPLICATION TO COMPACT OBJECTS}\label{s_appl}
The described GR polarized radiative transfer finds its application in accretion onto low luminosity active galactic nuclei (LLAGNs), in particular Sgr A*. The application of GR is necessary to infer the BH spin, which provides important information on the past evolution of the BH and the host galaxy itself. For example, the accretion efficiency in the AGN phase depends strongly on the value of BH spin \citep{shapiro}. The value of spin and spin orientation constraints the accretion and merger history \citep{rees}.

The detailed application to Sgr A* is reported in \citet{shcherbakov_penna}. Let us describe on the example of that paper, how one connects to observations. First, one constructs a set of dynamical models, preferably based on 3D GRMHD simulations. Then the GR polarized radiative transfer is performed for those models as described in the present work. The simulated spectral energy distributions (SEDs), linear (LP) and circular (CP) polarization fractions as functions of frequency $\nu_\infty$ are fitted to the observed quantities, representative for the quasi-quiescent state of accretion. The $\chi^2$ analysis is performed based on the inferred error or variability of the observed quantities. Then the best fits in the parameter space can be found. The probability density can be integrated over the full parameter space to obtain the most likely values and the confidence intervals of BH spin, inclination angle, position angle, and model parameters.

Let us now describe the effects, which lead to certain observed cyclo-synchrotron spectra, LP and CP fractions, and the electric vector position angle (EVPA) as functions of frequency. The effects are plenty, what proves it hard to disentangle and provide simple explanations of observations. It is in general challenging to achieve the realistic level of details in collisionless plasma modelling. The next step in \citet{shcherbakov_penna} might not be the most self-consistent. First, the radiation from LLAGNs appears to be variable in time. The simultaneous short observations can provide only the single snapshot of a system, not necessarily representative of a long term behavior. Thus, it is necessary to obtain the statistically significant sample of variability of both observed and simulated spectra to reliably estimate the average or typical flow parameters and BH spin. As recent research suggests \citep{dodds-eden}, the modelling of a single flare can successfully be done even without invoking GR.

As in the case of Sgr A*, the cyclo-synchrotron specific flux $F_\nu$ vs $\nu$ can have a peak. The peak frequency $\nu^*$ and flux $F_\nu^*$ do not necessarily correspond to the thermal cut-off of emission. Even a small percentage of the non-thermal electrons can radiate significantly more than the bulk of thermal electrons. For the efficient particle acceleration most of emission may come from the energetic electrons with cooling time $t_{\rm cool}$ about the time of inflow $t_{\rm in}$ from several BH gravitational radii $r_{\rm g}=G M/c^2.$ The gravitational redshift and Doppler shift due to relativistic motion can strongly modify the peak $\nu^*$ and $F_\nu^*.$

The LP fraction can provide constraints on flow density near the emitting region owing to beam depolarization effect. The LP fraction is the highest at high frequencies, where only a small region of the flow shines and beam depolarization is weak. As all regions of the flow radiate at lower frequencies, differential Faraday rotation and emission EVPA vary, the resultant LP fraction is subject to cancelations and quickly ceases with $\nu.$ However, cancelations at high $\nu$ may readily happen between two regions with similar fluxes and perpendicular EVPAs, those regions would have perpendicular magnetic fields.

The same change of EVPA with frequency can mimic the Faraday rotation. The finite rotation measure (RM)
\begin{equation}\label{RM_naive}
RM=\frac{EVPA_1-EVPA_2}{\lambda^2_1-\lambda^2_2}
\end{equation}
does not necessarily happen due to Faraday rotation $\sim \int n {\bmath B}\cdot {\bmath dl}$. Here $\lambda=c/\nu$ is the wavelength. In fact, the meaningful application of formula (\ref{RM_naive}) is limited to a toy case of cold plasma far from the emitting region with the homogeneous magnetic field. In reality, besides the change of emission EVPA with $\nu,$ the Faraday rotation coefficient $\rho_V (\nu)$ (see eq.\ref{ar_def}) is a function of frequency \citep{shcher_farad}. The differential rotation measure
\begin{equation}
dRM=dEVPA/d(\lambda^2)
\end{equation}
is the measured quantity \citep{marrone}. It should be used in constraining the models. Also, significant Faraday rotation can happen in the emitting region, what introduces the effect of differential optical depth. Thus, one can only fit the observed ${\rm EVPA}(\nu)$ and use it along with other observables to constrain the system free parameters. As $\rho_V (\nu)$ is a steeply declining function of temperature $\theta_e$ \citep{shcher_farad}, the relativistic charges contribute little to this quantity.

Substantial levels of circular polarization were recently found in Sgr A* (\citet{munoz}, in prep.). There are several effects producing finite CP. First recognized was the emissivity $\varepsilon_V$ in $V$ mode. According to \citet{melrose_emis} it only is a factor of $\gamma$ weaker than the total emissivity $\varepsilon_I.$ It produces the largest $V$ along the magnetic field. The Faraday conversion, transformation between the linear polarization and the circular, operates perpendicular to the magnetic field. The Faraday conversion coefficient $\rho_Q$ has a peculiar dependence on temperature of thermal plasma or particles' $\gamma$-factor \citep{shcher_farad}: $\rho_Q=0$ for cold plasma, $\rho_Q$ is exponentially inhibited for very hot plasma and reaches the maximum for transrelativistic plasma with $\theta_e\sim \gamma\sim {\rm several} \cdot m c^2/k_B.$ The exponential inhibition is an effect of finite ratio $\nu/\nu_B$ with peak $\rho_Q$ only around $\theta_e\sim10$ for $\nu/\nu_B\sim10^3.$ Thus, for hot bulk part of particle distribution with $\gamma\sim{\rm several}\cdot m c^2/k_B$ the non-thermal electrons do not contribute significantly to Faraday conversion. Note that this result supersedes \citet{ballantyne}, who following \citet{melrose97} neglected the importance of finite ratio $\nu/\nu_B.$ Similarly to linear polarization, the beam depolarization can lower the net value of circular polarization at low frequencies due to differential Faraday conversion.

In sum, there are often several explanations for the same observable quantity. One should not settle for a simplified model trying to reproduce the observations. Instead, the rigorous ray tracing and aposteriori explanations of a fit to the observables would be the preferred reliable way.

The proposed method has its limitation. The equation (\ref{fin_trans}) is valid for optically thick medium, but it fails to describe the behavior of a set of photons for Compton-thick medium. The encounters of photons with energetic electrons lead to significant changes in photon trajectory, whereas the previous discussion considered independent photons propagating along geodesics. Luckily, the synchrotron absorption cross-section in sub-mm is much larger than the Compton scattering cross-section, thus the optically thick medium near Sgr A* is Compton-thin and no modifications are needed for Sgr A*. However, a careful consideration of Compton scattering \citep{rybicki} is needed to describe the sub-mm spectrum of Compton-thick sources.

\section{DISCUSSION \& CONCLUSIONS}\label{s_discus}
In our endeavor to provide the complete and self-consistent description of GR polarized radiative transfer we conduct the full derivation starting from definitions and basic equations.
The goal is to make the easy, transparent, and error-free derivations, thus Mathematica 7 was used underway, the expressions were cross-checked. The absorptivities for thermal plasma were checked numerically against known synchrotron emissivities and cyclo-synchrotron approximations in \citet{sazonov,leung}. We stepped away from the standard textbooks and assumed "the opposite" observers' definition of circular polarization $V,$ carrying the definition through all other calculations. We chose the coordinate system with coplanar $\bmath k,$ $\tilde{\bmath e}^2,$ ${\bmath B}_0$ and derived the plasma response tensor $\tilde{\alpha}_{ik}$ in  ($\tilde{\bmath e}^1,$ $\tilde{\bmath e}^2,$ $\tilde{\bmath e}^3$) basis, also projecting it onto the transverse coordinates ${\bmath e}^1,$ ${\bmath e}^2.$ Repeating for completeness \citet{huangnew}, we tie the polarized radiative transfer equation in the latter coordinates (\ref{transfer}) with the transfer in $\bmath a,$ $\bmath b$ coordinates with the help of matrix (\ref{rmatrix}). The generalization of radiative transfer to GR is performed in the easiest way owing to the invariance of occupation numbers in different photon states and the invariance of the transformation between the states. Lorenz gauge (\ref{gauge}) helps to establish the correspondence between $4\times4$ and spatial $3\times3$ quantities and to correctly find the transverse vectors $\bmath a,$ $\bmath b$ in the co-moving locally-flat reference frame. The transformation from the lab frame with Kerr metric to that frame is explicitly given. The intricacies of application of GR polarized radiative transfer to LLAGNs are discussed. As the transfer incorporates many physical effects, a priori guessing of the most important effects is discouraged in favor of full calculation. The provided interface of dynamical models and observations is waiting for its applications.

The treatment of particles distributions is still limited. In the current state the calculations are optimized for isotropic in pitch-angle distributions and become especially simple for thermal particle distribution. The integration over the pitch-angle in formula (\ref{resp_def}) is in general impossible to perform for non-isotropic distributions. In this case, the integral over $\xi$ should be done first analytically. This calculation is left for future work.
\section*{ACKNOWLEDGMENTS} The authors are grateful to Charles Gammie, James Moran, Diego Munoz, and Ramesh Narayan for fruitful discussions, Akshay Kulkarni and Robert Penna for help with the transformation to the locally-flat co-moving frame, and to anonymous referee for helpful comments. Charles Gammie kindly provided us a draft of their paper (\citet{gammie}, in prep.) on covariant formalism of polarized ray tracing.

The work is partially supported by NASA grant NNX08AX04H to RVS and China Postdoctoral Science Foundation grant 20090450822 to LH.

\label{lastpage}
\end{document}